\newcommand{\dia}{\!\!\!\!\!\!\!\not\,\,\,\,}
\newcommand{\nn}{\nonumber}

\newcommand{\be}{\begin{eqnarray}}
\newcommand{\ee}{\end{eqnarray}}

\documentclass[twocolumn,aps,showpacs]{revtex4}
\usepackage{graphics}
\begin{document}
\title{Dynamical mass generation for fermions in quenched Quantum
Electrodynamics at finite temperature}
\author{Alejandro Ayala$^\dagger$, Adnan Bashir$^\ddagger$ } 
\affiliation{$\dagger$Instituto de Ciencias Nucleares, Universidad Nacional 
Aut\'onoma de M\'exico, Apartado Postal 70-543, M\'exico Distrito Federal 
04510, M\'exico \\
$^\ddagger$Instituto de F{\'\i}sica y Matem\'aticas,
Universidad Michoacana de San Nicol\'as de Hidalgo, Apartado Postal
2-82, Morelia, Michoac\'an 58040, M\'exico}

\begin{abstract} 
We study the dynamical generation of masses for fundamental fermions in
quenched quantum electrodynamics (qQED) at finite temperature in the bare
vertex approximation, using Schwinger-Dyson equations (SDE). Motivated by
perturbation theory, a further simplification is introduced by taking
the wave function renormalization to be unity. In the zeroth mode
approximation, the SDE for the fermion propagator resembles 
QED in 2+1 dimensions (QED3) at zero temperature with an effective
dimensionful coupling $\alpha'=\alpha T$. For a fixed temperature,
mass is dynamically generated above a certain critical value of this
coupling. As expected, raising the temperature restores chiral
symmetry and fermions become massless again. We also argue that by
summing over the frequency modes and under suitable simplifications,
qualitative aspects of the result do not undergo significant changes. 
\end{abstract}



\maketitle

\section{Introduction}

The perturbative treatment of gauge field theories at zero temperature is 
valid only for weak coupling, which fails to explain certain
non-perturbative aspects of these theories such as dynamical mass
generation for fermions. Schwinger-Dyson equations (SDEs) provide us
with a natural platform to study this phenomenon because their
derivation makes no assumption about the strength of the
interaction. In quantum electrodynamics (QED) at
zero temperature, it is well known that above a certain critical value
of the coupling, fermions can acquire masses through self interaction without
the necessity of a non zero bare mass. This fact is in marked contrast 
to perturbation theory which prohibits the appearance of masses in
this fashion as long as the bare mass is zero even if we did an
all order resummation using renormalization group equations.
In a thermal background, temperature is another parameter which comes 
into play and we expect criticality in it too, triggering mass generation
below a certain temperature and restoring chiral symmetry if we go
above this value. The relevant SDE to investigate such a behavior is
that of the fermion propagator. 

SDEs are an infinite tower of coupled equations relating various
Green's functions to each other. The two-point Green's function of the
fermion propagator is coupled to all higher order Green's
functions. Therefore, a non-perturbative truncation is essential to
convert it into a solvable form. In quenched quantum electrodynamics (qQED), 
a favorite starting
point is to make an  ansatz for the fermion-photon vertex and then
study the fermion propagator equation in its decoupled form.
Inspired by some key features of gauge theories such as Ward-Green-Takasahi
identities, Landau-Khalatnikov-Fradkin transformations, multiplicative
renormalizability of the fermion propagator, etc.,   
significant advances have been made in the hunt for a reliable vertex ansatz
at zero temperature and the corresponding study of dynamical chiral
symmetry breaking. However, for finite temperature, this venue has
been much less explored, partially because of the richer tensor
structure of the propagator and the vertex
involved~\cite{AB,Kondo,Harada,Fukazawa,Ikeda,Fueki1,Fueki}. The case
of three dimensional QED (QED3)
at finite temperature has been studied in relatively more detail in
the bare vertex
approximation~\cite{{Dorey1},{Dorey2},{Aitchison2},{Aitchison3}, 
{Triantaphyllou1},{Triantaphyllou2},{Lee1}}.

In this paper, we undertake the study of QED4
in the quenched approximation at finite temperature. We study the
corresponding Schwinger-Dyson equation for the fermion propagator in the
bare vertex approximation. Under the additional assumptions of setting the  
wave function renormalization to unity and picking out only the most
dominant frequency mode, the theory resembles QED3 at zero temperature (see
also \cite{Ginsparg}). The effective coupling $\alpha T$ is
dimensionful. In order to pursue the resemblance of the theory with
QED3, we can take $\alpha'=\alpha T$ along with $T$ as independent
parameters. The only notable difference is a temperature dependent 
mass term. The effect of this term is such that keeping the
temperature constant, the increase of the coupling beyond a certain
critical value makes the fermion to acquire a mass. Conversely,
increasing the temperature above a threshold value, we observe chiral
symmetry restoration. We also confirm this behavior
analytically. Moreover, increasing the value of the coupling, the
critical value of the temperature also gets multiplied by the same
factor. We must emphasize that the existence of criticality for the 
coupling and temperature is in contrast to the behavior of QED3 at
zero temperature where no criticality exists: if masses are generated 
for one value of the coupling, they are generated for all other values of
the coupling. Although not explicitly considered, the analysis
can easily be extended to the case in which $\alpha$ and $T$ are
taken as the independent parameters.

The work is organized as follows: In Sec.~\ref{II} we set up
the SDE for the mass function in qQED with bare vertex at
finite temperature using the imaginary-time formalism, starting from a
simplified tensor structure for the fermion propagator. In
Sec.~\ref{III}, we work under the assumption of considering only the
contribution from the lowest Matsubara frequency finding a non-trivial
numerical solution for the mass function. We study critical behavior
of the mass function with the coupling strength as well as with
temperature. With further simplifications in the integral equation, 
we also study this critical behavior analytically, confirming the
numerical results. In Sec.~\ref{sum} we relax the assumption that
considered the lowest Matsubara frequency as the single contributing
mode and perform the sum over frequencies. We analytically study the
resulting integral equation under the same simplifying assumptions as
before and argue that, except for slightly different numerical
coefficients, the behavior of the mass function appears to be
qualitatively the same. We finally summarize and discuss our results
in Sec.~\ref{concl}. 

\section{The Fermion Propagator and the Mass Function}\label{II}

\noindent
The Minkowski space  Feynman rules for the QED bare fermion propagator, photon 
propagator and fermion-photon vertex are, respectively, 
\be
   i S_F^{0}(P)&=&\frac{i}{\not \! P}\;,\nonumber\\
   D_0^{\mu \nu}(Q)&=&\frac{i}{Q^2} 
   \left[ - g^{\mu \nu} + (1-\xi)
   \frac{ Q^{\mu}Q^{\nu}}{Q^2}\right]\;,\nonumber\\ 
   i g \Gamma^{\mu}_0(K,P)&=&i g \gamma^{\mu}\;,
   \label{bareprop}
\ee 
(hereafter we use capital letters to refer to four-vectors, whereas 
lower case letters are used for their components) where $g$ is the
electromagnetic coupling and $\xi$ is the usual covariant gauge
parameter. Making use of the above rules, the SDE for the fermion
propagator can be written as 
\be
   S_F^{-1}(P)\!\!&=&\!\! (S_F^0)^{-1}(P)\nonumber\\
   \!\!&-&\!\! ig^2 \int \frac{d^4K}{(2\pi)^4} \Gamma^{\mu}(K,P) 
   S_F(K) \gamma^{\nu} D^0_{\mu \nu}(Q)\, ,
   \label{prop}
\ee
where $Q=K-P$, $\Gamma^{\nu}(K,P)$ is
the full fermion-photon vertex and $S_F(P)$ is the full fermion
propagator.

At finite temperature, the most general tensor form for the
fermion propagator can be obtained by noticing that, in addition to
$1$ and $\not \!\!P\;$, there are also two other Lorentz structures,
namely $\not \! U$ and $\not \! P\not \! U$. Therefore, 
we can write
\be
   S_F(P)=\frac{F}{P\dia \, - \, b \, U\dia \, 
   - \, c \, P\dia \, U\dia \, - \,  {M}}\, .
   \label{ferprop}
\ee
The Lorentz invariant functions $F$, $b$, $c$ and ${M}$ will
in general depend on two Lorentz scalars $P^2$ and $U \cdot P$,
where $U^\mu$ is the four-velocity of the heat bath as seen from a
general frame. We can choose these scalars to be  
\be
   p_0&\equiv & P^\mu U_\mu\nonumber\\
   p&\equiv &[(P^\mu U_\mu )^2-P^2]^{1/2} \,.
   \label{escalars}
\ee
The functions $F$ and ${M}$ are called the wave function
renormalization and mass function, respectively. Notice that for a
parity conserving theory such as QED, $c=0$~\cite{Weldon}. 

For our purposes, we will work in the approximation where $b=0$ and
$F=1$. The justification for this approximation stems from the
fact that perturbatively $F=1 + {\cal O}(\alpha)$ and $b =
{\cal O}(\alpha)$. If $\alpha$ is small, we naturally expect 
${F}\approx 1$ and $b\approx 0$. Though it is well known that
at zero temperature, inclusion of ${F}$ is essential to recover
the gauge invariance of the results such as the generated fermion
mass~\cite{BP1}, it is also known that when ignoring ${F}$, the
qualitative results, such as the shape of the fermion mass as a
function of the coupling strength, are not changed. At finite
temperature, we can expect that ${F}$ and $b$ play a similar role
for the restoration of the gauge
invariance~\cite{Aitchison2}. Nevertheless, to keep the discussion as
simple as possible, here we will write the fermion propagator as
\be
   S_F(P)=\frac{1}{\not \! P- { M}(P)} \;,
   \label{fullprop}
\ee
ignoring ${F}$ and $b$ in Eq.~(\ref{ferprop}).

Taking the trace of Eq.~(\ref{prop}), making use of 
Eqs.~(\ref{bareprop}) and~(\ref{fullprop}), and
replacing the full vertex by its bare counterpart, the equation
for the mass function can be written in Minkowski space as
\begin{eqnarray}
   {M}(P)= -i\frac{(3+\xi)\alpha}{4 \pi^3} \int d^4K
   \, \frac{{M}(K)}{K^2-{M}^2(K)} \, \frac{1}{Q^2}\;.
   \label{massfuncmink}
\end{eqnarray}
We work explicitly in the imaginary-time
formulation of thermal field theory, taking $K=(i\omega_n,{\bf k})$,
$P=(i\omega,{\bf p})$ where $\omega_n=(2n+1)\pi T$ and
$\omega=(2m+1)\pi T$ are discrete fermion frequencies, with $T$
being the temperature and $n$, $m$ integers. In this formalism,
Eq.~(\ref{massfuncmink}) becomes
\begin{eqnarray}
   {M}(P)&=&\frac{(3+\xi)\alpha}{2 \pi^2}
   T \sum_n\int d^3k\Big\{
   \frac{{M}(K)}{[\omega_n^2+{\bf k}^2+{M}^2(K)]}\nonumber\\
   &\times&
   \frac{1}{[(\omega_n-\omega)^2 + |{\bf k}-{\bf p}|^2]}\Big\}\;.
   \label{massfunceucl}
\end{eqnarray}

\section{Zeroth mode approximation}\label{III}

\subsection{The full equation}

\begin{figure}[t]     
\rotatebox{-90}                     
{\resizebox{6cm}{7.8cm}{\includegraphics{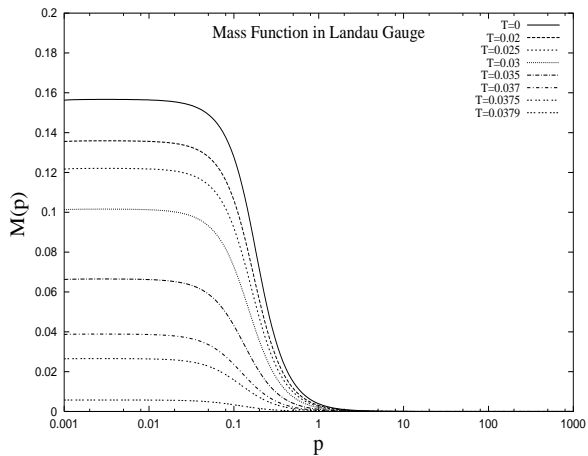}}}  
\vspace{2mm}                 
\caption{The mass function ${M}(p)$ for various values of temperature
and $\alpha'=1/4 \pi$ in the Landau gauge.} 
\label{MvsT} 
\end{figure}

A trivial solution to Eq.~(\ref{massfunceucl}) is ${M}(P)=0$,
which corresponds to the usual perturbative solution. However, we are
interested in a non-trivial solution. To pursue it, as a
first step we make the approximation where
we take only the contribution of the zeroth mode, i.e., $n=0$, in
the sum over frequencies in Eq.~(\ref{massfunceucl}) and
correspondingly take the same value for $\omega$. We relax this
approximation in the following section. It appears natural to take 
$\alpha T=\alpha'$ and $T$ as independent variables. After carrying out 
the angular integration, Eq.~(\ref{massfunceucl}) becomes 
\be
   {M}(p) = \frac{(3+\xi )\alpha'}{2 \pi p}
   \int_0^{\infty}\frac{dk \, k {M}(k)}{\pi^2 T^2 + k^2 + {M}^2(k)}
   {\rm ln} \left|\frac{k+p}{k-p}\right| ,  
   \label{nosum}
\ee
where $p$ and $k$ are the magnitudes of the vectors ${\bf p}$ and
${\bf k}$, respectively. We solve Eq.~(\ref{nosum}) by converting the
integral equation into a set of simultaneous non-linear equations. The
range of integration used is $k=10^{-3}$ to $10^{3}$. The number of
points per decade to carry out the integration was taken to be 31
which was found to be a sufficiently high number to obtain the
accuracy with which the results have been quoted. In
Fig.~\ref{MvsT}, we plot the mass function $M$ for $\alpha'=1/4 \pi$
and various values of $T$ in the Landau gauge, i.e., $\xi=0$. With the
increase of temperature, the generated mass decreases rapidly. The
large $p^2$ behavior of the generated mass is depicted in Table
I. We find that the mass function decreases as $1/p^2$ in the
asymptotic region of momentum.

\begin{center}
\begin{figure}[ht]     
\rotatebox{-90}                     
{\resizebox{6cm}{7.8cm}{\includegraphics{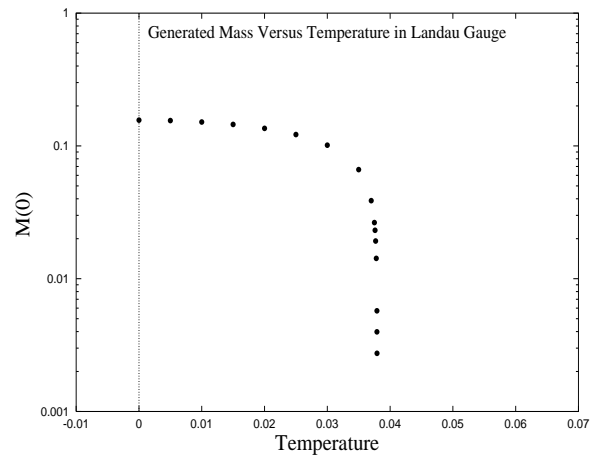}}}             
\caption{The approximated generated mass ${M}(0)$ for the fermion for
various values of temperature and $\alpha'=1/4 \pi$ in the Landau gauge.} 
\label{M0vsT}                                        
\end{figure}          
\end{center}

The generated mass can be approximated by $M(0)$~\cite{BHR1,Atkinson},
since for small $p$, $M(p)$ is rather constant. We draw it as a
function of temperature in Fig.~\ref{M0vsT} in the Landau gauge. One
can see a clear indication 
of the existence of a critical value $T_c$ of temperature above which there is 
no dynamical generation of mass. $T_c$ was found to be around 0.037915.
In order to see whether this critical value of the coupling can be obtained
analytically, we analyze an approximated version of Eq.~(\ref{nosum}) in
section \ref{C}.

\subsection{Comparison with QED3 at zero temperature}

It is worth mentioning 
that at $T=0$, Eq.~(\ref{nosum}) is the same as the one in QED3 at
zero temperature in the approximations of using the bare counterpart of the
full vertex and setting $F=1$ (see Eq. (2.7) of Ref. 
\cite{BHR1}), with
the only difference that the factor $2+\xi$ has been replaced with $3+\xi$.
The origin of this only difference is the the fact that the trace algebra is
carried out in different dimensions in the two cases. 
 We now mention an important contrast with QED3 at zero temperature where
under identical assumptions, equation for the mass function reads as
\be
   {M}(p) = \frac{(2+\xi )\alpha}{\pi p}
   \int_0^{\infty}dk\frac{k {M}(k)}{k^2 + {M}^2(k)}
   {\rm ln} \left|\frac{k+p}{k-p}\right|\, ,
   \label{masszeroT}
\ee
This equation is scale invariant. The solution for an arbitrary value of 
$e^2$ can be obtained from $e^2=1$ in the following fashion:
\be
       M(p,e^2) = e^2 M(p/e^2,1)  \;. \nn
\ee
Therefore, if chiral symmetry breaking solution exists for one value of 
coupling it exists for all values of coupling. There is no critical value 
of $e^2$ or $\alpha$. A mathematically similar scaling law at finite
temperature is
\be
     M(p,e^2,T) = e^2 M(p/e^2,1,T/e^2) \;.  \nn
\ee
We have seen that for $e^2=1$, there is a critical temperature,
$T_c=0.037915$, which separates the chirally asymmetric solution from the
symmetric one, in the Landau gauge. Therefore, we have
\be
      M(p,1,T_{>c})=0      \hspace{15mm} \forall \hspace{5mm} p  \nn
\ee
for all temperatures $T_{>c}$ above its critical value. What happens if we
double the value of coupling $e^2$? The scaling law implies that
\be
    M(p,2,T) = 2 M(p/2,1,T/2)   \;. \nn
\ee

\begin{tabular}{|crr|}  \hline 
&     &    \\
\multicolumn{1}{|c}{$T$}  &
\multicolumn{1}{c}{$p$} &
\multicolumn{1}{c|}{$p^2\,M(p^2)$}  \\
&     &    \\                                 \hline  \hline
&     &    \\
0.0   &  1000      &  3.89920     \\   
      &  640.4     &  3.89940     \\   
      &  304.699   &  3.89947     \\   
&     &    \\                        \hline
&     &    \\
0.005 &  1000      &  3.84612     \\   
      &  640.4     &  3.84632     \\   
      &  304.699   &  3.84638     \\   
&     &    \\                        \hline
&     &    \\
0.01  &  1000      &  3.68662     \\   
      &  640.4     &  3.68681     \\   
      &  304.699   &  3.68687     \\   
&     &    \\                        \hline
&     &    \\
0.015 &  1000      &  3.41973     \\   
      &  640.4     &  3.41990     \\   
      &  304.699   &  3.41996     \\   
&     &    \\                        \hline
&     &    \\
0.02  &  1000      &  3.04294     \\   
      &  640.4     &  3.04310     \\   
      &  304.699   &  3.04315     \\   
&     &    \\                        \hline
&     &    \\
0.025 &  1000      &  2.54986     \\   
      &  640.4     &  2.54999     \\   
      &  304.699   &  2.55003     \\ 
&     &    \\                        \hline 
&     &    \\
0.03  &  1000      &  1.92211     \\   
      &  640.4     &  1.92220     \\   
      &  304.699   &  1.92224     \\
&     &    \\                        \hline
&     &    \\
0.035 &  1000      &  1.08292     \\   
      &  640.4     &  1.08297     \\   
      &  304.699   &  1.08299     \\        
  &     &    \\                                \hline
\end{tabular}
\begin{tabular}{|crr|}  \hline 
&     &    \\
\multicolumn{1}{|c}{$T$}  &
\multicolumn{1}{c}{$p$} &
\multicolumn{1}{c|}{$\,p^2\,M(p^2)$}  \\
&         &    \\                        \hline \hline
&         &    \\
0.037     &  1000      &  0.579892    \\   
          &  640.4     &  0.579921    \\   
          &  304.699   &  0.579931    \\   
&         &    \\                        \hline
&         &    \\
0.0375    &  1000      &  0.386170     \\   
          &  640.4     &  0.386189     \\   
          &  304.699   &  0.386196     \\   
&         &    \\                        \hline
&         &    \\
0.0376    &  1000      &  0.336004     \\  
          &  640.4     &  0.336021     \\  
          &  304.699   &  0.336027     \\   
&         &    \\                        \hline
&         &    \\
0.0377    &  1000      &  0.277625     \\  
          &  640.4     &  0.277640     \\   
          &  304.699   &  0.277644     \\  
&         &    \\                        \hline
&         &    \\
0.0378    &  1000      &  0.204152     \\  
          &  640.4     &  0.204163     \\   
          &  304.699   &  0.204166     \\  
&         &    \\                        \hline
&         &    \\
0.0379    &  1000      &  0.0818793     \\  
          &  640.4     &  0.0818835     \\   
          &  304.699   &  0.0818849     \\  
&         &    \\                        \hline
&         &    \\
0.03791   &  1000      &  0.0568280     \\  
          &  640.4     &  0.0568309     \\   
          &  304.699   &  0.0568319     \\
&         &    \\                        \hline
&         &    \\
0.037915  &  1000      &  0.0391043     \\  
          &  640.4     &  0.0391063     \\   
          &  304.699   &  0.0391070     \\
  &     &    \\                               \hline   
\end{tabular}  \label{table1} \\ \\
{TABLE~1. The large $p^2$ behavior of ${M}(p)$ 
for various values of temperature}   \\

\begin{tabular}{|crr|}  \hline 
&     &    \\
\multicolumn{1}{|c}{$T$}  &
\multicolumn{1}{c}{$p$} &
\multicolumn{1}{c|}{$p^2\,M(p^2)$}  \\
&     &    \\                                 \hline  \hline
&     &    \\
0.0   &  1000      &  2.41264            \\   
      &  640.4     &  2.41257     \\   
      &  304.699   &  2.41219     \\   
&     &    \\                        \hline
&     &    \\
0.005 &  1000      &  2.35491     \\   
      &  640.4     &  2.35485     \\   
      &  304.699   &  2.35448     \\   
&     &    \\                        \hline
&     &    \\
0.01  &  1000      &  2.18138     \\   
      &  640.4     &  2.18132     \\   
      &  304.699   &  2.18098     \\   
&     &    \\                        \hline
&     &    \\
0.015 &  1000      &  1.89032     \\   
      &  640.4     &  1.89027     \\   
      &  304.699   &  1.88997     \\   
&     &    \\                        \hline
&     &    \\
0.02  &  1000      &  1.47509     \\   
      &  640.4     &  1.47505     \\   
      &  304.699   &  1.47482     \\   
&     &    \\                        \hline
&     &    \\
0.025 &  1000      &  0.905525     \\   
      &  640.4     &  0.905501     \\   
      &  304.699   &  0.905357     \\ 
&     &    \\                        \hline 
&     &    \\
0.026 &  1000      &  0.762463     \\   
      &  640.4     &  0.762443     \\   
      &  304.699   &  0.762322     \\
&     &    \\                        \hline
&     &    \\
0.027 &  1000      &  0.600565     \\   
      &  640.4     &  0.600549     \\   
      &  304.699   &  0.600454     \\        
  &     &    \\                               \hline   
\end{tabular}
\begin{tabular}{|crr|}  \hline 
&     &    \\
\multicolumn{1}{|c}{$T$}  &
\multicolumn{1}{c}{$p$} &
\multicolumn{1}{c|}{$\,p^2\,M(p^2)$}  \\
&         &    \\                        \hline \hline
&         &    \\
0.028     &  1000      &  0.402201    \\   
          &  640.4     &  0.402190    \\   
          &  304.699   &  0.402126    \\   
&         &    \\                        \hline
&         &    \\
0.0285    &  1000      &  0.268127     \\   
          &  640.4     &  0.268120     \\   
          &  304.699   &  0.268078     \\   
&         &    \\                        \hline
&         &    \\
0.0286    &  1000      &  0.234034     \\  
          &  640.4     &  0.234028     \\  
          &  304.699   &  0.233991     \\   
&         &    \\                        \hline
&         &    \\
0.0287    &  1000      &  0.194810     \\  
          &  640.4     &  0.194805     \\   
          &  304.699   &  0.194774     \\  
&         &    \\                        \hline
&         &    \\
0.0288    &  1000      &  0.146417     \\  
          &  640.4     &  0.146413     \\   
          &  304.699   &  0.146390     \\  
&         &    \\                        \hline
&         &    \\
0.0289    &  1000      &  0.0724110     \\  
          &  640.4     &  0.0724091     \\   
          &  304.699   &  0.0723976     \\  
&         &    \\                        \hline
&         &    \\
0.02891   &  1000      &  0.0602952     \\  
          &  640.4     &  0.0602936     \\   
          &  304.699   &  0.0602841     \\
&         &    \\                        \hline
&         &    \\
0.02892   &  1000      &  0.0451881     \\  
          &  640.4     &  0.0451869     \\   
          &  304.699   &  0.0451798     \\
&         &    \\                        \hline
\end{tabular}  \label{table2} \\ \\ \\
{TABLE~2. The large $p^2$ behavior of ${M}(p)$ 
for various values of temperature for the approximated log.}
\vspace{1cm}

Let us substitute $T \rightarrow 2T_c$ in this identity which gives
\be
      M(p,2,2T_c) = 2 M(p/2,1,T_c) \;. \nn
\ee
Thus $M(p,2,2T_{>c})=0$, i.e., the value of the critical coupling gets
doubled by doubling the value of coupling $e^2$. To confirm this scaling
law we study Eq.~(\ref{nosum})  for $\alpha'=1/2 \pi$ and find that the new 
value of critical temperature is indeed roughly $0.075825$. This is
shown in Fig. 3.
\begin{figure}[t]     
\rotatebox{-90}                     
{\resizebox{6cm}{7.8cm}{\includegraphics{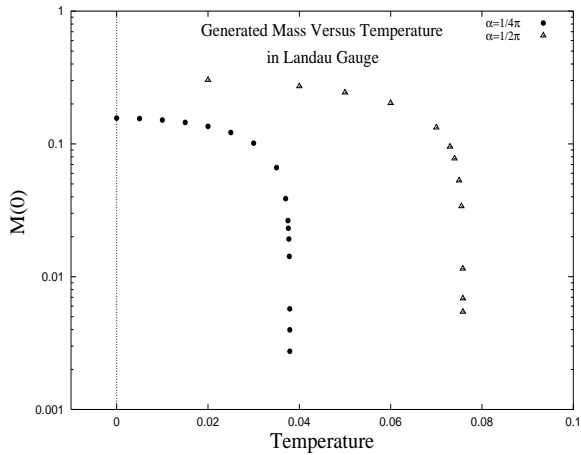}}}  
\vspace{2mm}                 
\caption{The approximated generated mass ${M}(0)$ for the fermion for
various values of temperature and two different values of
coupling, $\alpha'=1/4 \pi$ and $\alpha'=1/2 \pi$ in the Landau gauge.} 
\label{Mvsalpha} 
\end{figure}          

\subsection{The equation with approximated log}\label{C}

\begin{center}
\begin{figure}[t]     
\rotatebox{-90}                     
{\resizebox{6cm}{7.8cm}{\includegraphics{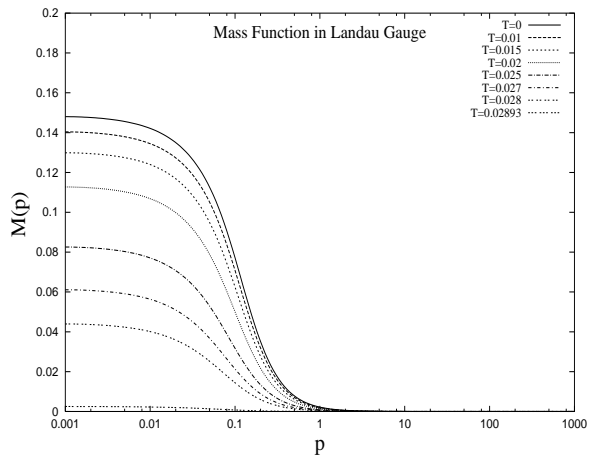}}}                
\caption{The mass function ${M}(p)$ for various values of temperature
and $\alpha=1/4 \pi$ for the approximated logarithm in the Landau gauge.} 
\label{applog1} 
\end{figure}          
\end{center}

We can simplify Eq.~(\ref{nosum}) by employing the approximation \cite{BR1},
\be
  {\rm ln}\left|\frac{x+y}{x-y}\right| &\equiv& \frac{2y}{x} \theta(x-y) 
+ \frac{2x}{y} \theta(y-x)  \;,
\label{logapprox}
\ee
which is a good approximation for $x<<y$ and $x>>y$. Moreover, for a small
generated mass, we can also neglect the factor $M^2(k)$ in the denominator.
Equation~(\ref{nosum}) can then be written as
\be
   {M}(p)\!\!&=&\!\! \frac{(3+\xi )\alpha'}{\pi}  
\int_0^{\infty}\! dk \frac{ {M}(k)}{\pi^2 T^2 + k^2}
     \Big[ \theta(k-p)  \nn  \\
&& \hspace{35mm} + \frac{k^2}{p^2} \theta(p-k)   \Big]. 
  \label{applog}
\ee
If we repeat the numerical analysis of the previous sub-section, we find that
the qualitative results do not undergo any drastic change. The mass function 
still lowers its height with temperature and above a critical value of
the temperature (now 0.02893, see Fig. 5), only the trivial solution
exists, i.e., 
$M(p)=0$. This behavior is shown in Fig.~\ref{applog1} in the Landau
gauge. Moreover,
the large momentum behavior of the mass function is again $1/p^2$ as
shown in Table II. This behavior for asymptotic values of momentum can be
confirmed by converting Eq.~(\ref{applog}) into a second order ordinary
differential equation 
\be
    \frac{d}{dp}\left[ p^3 \frac{dM(p)}{dp} \right] + 
\frac{4(3+\xi) \alpha' M(p)}{\pi}  = 0 \;,
\label{diflog}
\ee
with boundary conditions
\be
p^3 \frac{dM(p)}{dp}=0 \Bigg|_{p=0} \hspace{10mm} {\rm and} \hspace{10mm} 
M(p)=0 \Big|_{p \rightarrow \infty} \;,
\nonumber
\ee
where we have dropped $\pi T$ in comparison with $p$. 
The solution of this equation is
\be
M(p) &=& \frac{8(3+\xi )\alpha'}{\pi p} \Bigg[ 
c_1 J_2\left(4 \sqrt{\frac{(3+\xi )\alpha'}{\pi p}}\ \right) \nn \\
&& \hspace{17mm} - 
c_2 Y_2\left(4 \sqrt{\frac{(3+\xi )\alpha'}{\pi
    p}}\ \right) \Bigg] \;,
\ee 
where $J(x)$ and $Y(x)$ are the Bessel functions of the first and
second kind, respectively. The second boundary condition imposes
$c_2=0$. As $J(1/\sqrt{p}) \rightarrow 1/p$ when $p \rightarrow \infty$,
we conclude that $M(p) \rightarrow 1/p^2$ in this limit, as found 
numerically. However, such
an analytic treatment washes out the information about the effect of 
temperature and we cannot find the critical value of coupling. To retain 
the effect of temperature, note that the factor $\pi T$ in
Eq.~(\ref{applog}) serves as an infrared cutoff and therefore, we
can rewrite this equation as
\be
   {M}(p) &=& \frac{(3+\xi )\alpha'}{\pi}  
\int_{\pi T}^{\infty}dk\frac{ {M}(k)}{k^2}
     \Big[ \theta(k-p)  \nn  \\ 
&& \hspace{35mm} + \frac{k^2}{p^2} \theta(p-k)   \Big] \;. 
   \label{applogsim}
\ee
The corresponding differential equation is exactly as we had obtained before.
Only the first boundary condition changes to
\be
     p^3 \frac{dM(p)}{dp} \Bigg|_{p=\pi T} &=& 0 \;.
\ee
This boundary condition leads us to the following equation at the critical
coupling
\be
    J_2(4a) + a \left[J_1(4a)-J_3(4a)  \right] &=& 0  \;, \label{critical}
\ee
where 
\be
a=\sqrt{(3+\xi )\alpha'/ \pi^2 T}  \label{analytical}\;. 
\ee
Note that $a$ being a
constant [determined 
by Eq.~(\ref{critical})], we see that at the critical coupling, $\alpha'$ and
$T$ are proportional to each other, i.e., if we increase $\alpha'$, the
critical value of coupling will increase by the same factor. The first
non-trivial zero of Eq.~(\ref{critical}) yields $a=0.958$ for 
$\alpha'=1/4 \pi$.
This implies $T=0.02636$ in the Landau gauge, which is
within $10\%$ of the actual numerical value. 

     What happens when we are in a gauge other than the Landau gauge?
In principle, the critical values $\alpha_c$ and $T_c$ should be gauge
independent because they trigger a phase transition. However, we do not 
expect it to be the case in the present work primarily because of the 
following reasons~: (1) We assume the wave function renormalization 
$F(p)$ to be 1. The studies at zero temperature indicate that the inclusion of
the equation for $F(p)$ plays a crucial role in the partial restoration of
gauge invariance, \cite{BHR1}. (2) The rainbow
approximation neither respects the Ward-Green-Takahashi identity
nor the Landau-Khalatnikov-Fradkin transformations, both being the 
consequences of gauge invariance. The gauge dependence of $T_c$ as a result of
the numerical and analytical analyses  has been 
plotted in  Fig.~(\ref{gauge}). As another confirmation of the analytical
result, Eq.~(\ref{analytical}), we find that in our numerical calculation,
the critical temperature in an arbitrary covariant gauge, $T_c(\xi)$, is 
obtained from the one in the Landau gauge through the relation 
$T_c(\xi)=[(3+\xi)/3] T_c$.

\begin{center}
\begin{figure}[t]     
\rotatebox{-90}                     
{\resizebox{6cm}{7.8cm}{\includegraphics{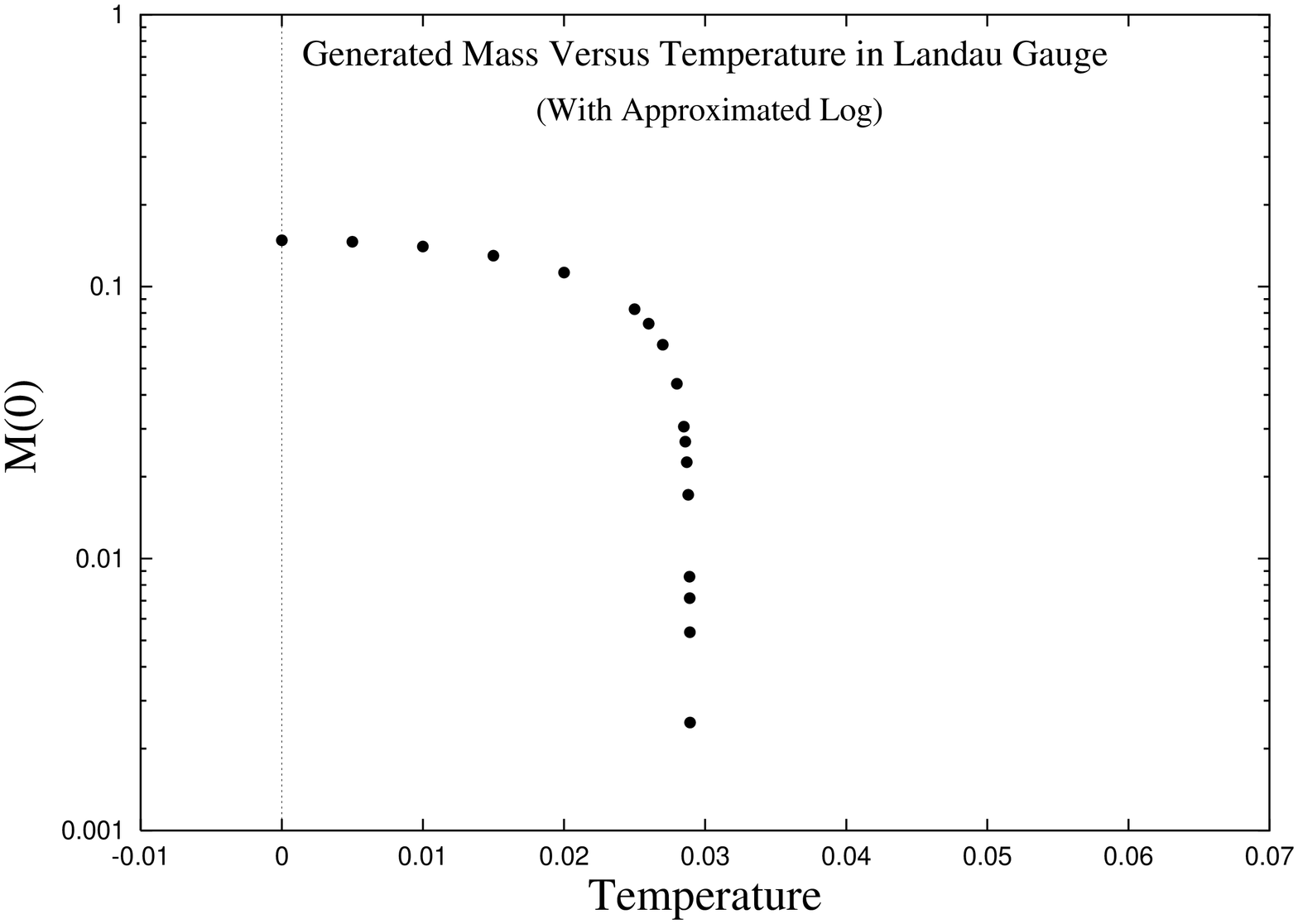}}}  
\vspace{2mm}                 
\caption{The approximated generated mass ${M}(0)$ for the fermion for
various values of temperature and $\alpha=1/4 \pi$ for the approximated
logarithm in the Landau gauge.} 
\label{fnoWTI} 
\end{figure}          
\end{center}

\begin{center}
\begin{figure}[t]     
\rotatebox{-90}                     
{\resizebox{6cm}{7.8cm}{\includegraphics{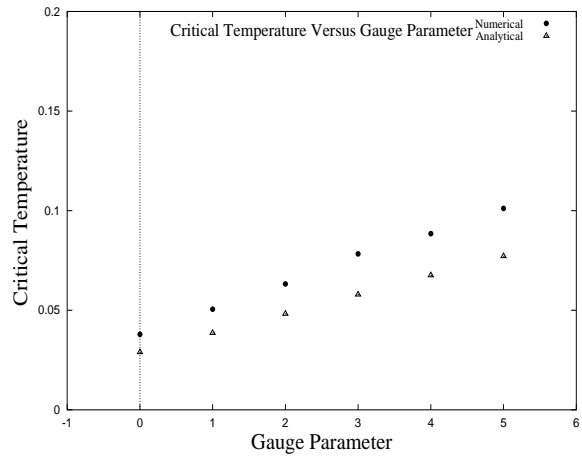}}}  
\vspace{2mm}                 
\caption{Critical temperature $T_c$ versus the gauge parameter $\xi$. The
upper and lower curves correspond to numerical and analytical results,
respectively.} 
\label{gauge} 
\end{figure}          
\end{center}

\section{Summing over Matsubara frequencies}\label{sum}

Here we want to show that, under some simplifications to
analytically handle the expressions, the results obtained in the
previous section appear to remain qualitatively unchanged, when
one considers the 
contribution from Matsubara frequencies other than the lowest one. Let
us stress that this point is by no means evident since there are known
examples in QED for which the contribution from the zeroth mode is exactly
canceled by the contribution of the rest of the modes. One such
example is provided by the one-loop fermion damping rate at finite
temperature, which, as expected on physical grounds, when evaluated on
the free mass shell vanishes due to the absence of phase space. This
result can also be seen to arise as a consequence of the 
exact cancellation between the (divergent) contribution from the
zeroth Matsubara mode and the rest of the modes~\cite{Blaizot}.

Since we will be primarily interested in order of magnitude estimates,
we will carry out some rough approximations aimed to preserve the
possibility to analytically extract the behavior of the mass function
in the critical region. 
The starting point is the sum over Matsubara frequencies indicated in
Eq.~(\ref{massfunceucl}) which we rewrite here as
\be
   I(i\omega, {\bf p})&\equiv& T \sum_n{M}(K)\nonumber\\
   &\times&\frac{1}{[\omega_n^2+{\bf k}^2+{M}^2(K)]
   [(\omega_n-\omega)^2 + |{\bf k}-{\bf p}|^2]}\, .\nonumber\\
   \label{explsum}
\ee

\vspace{0.5cm}
As a first approximation, we neglect a possible dependence of ${M}(K)$ 
on the frequency, writing ${M}(K)={M}(k)$. In this
approximation, the sum in Eq.~(\ref{explsum}) can be computed using
standard methods~\cite{LeBellac} with the result
\be
   I(i\omega, {\bf p})\!&=&\! -\frac{1}{4E_1E_2}\Big\{
   \left[ 1 + f(E_2) - \tilde{f}(E_1) \right]\nonumber\\
   \!&\times&\! 
   \left( \frac{1}{i\omega - E_1 - E_2} - \frac{1}{i\omega + E_1 +
   E_2}\right)\nonumber\\
   \!&+&\! \left[ f(E_2) + \tilde{f}(E_1) \right]\nonumber\\
   &\times&
   \left( \frac{1}{i\omega - E_1 + E_2} - \frac{1}{i\omega + E_1 -
   E_2}\right)\Big\}\! \;,\nonumber\\
   \label{sumres}
\ee
where $E_1=\sqrt{k^2+{M}^2(k)}$, $E_2=|{\bf k}-{\bf p}|$ and
\be
   f(x)&=&\frac{1}{e^{x/T}-1}\nonumber\\
   \tilde{f}(x)&=&\frac{1}{e^{x/T}+1}
   \label{BEFD}
\ee
are the Bose-Einstein and Fermi-Dirac distributions,
respectively. Equation~(\ref{sumres}) contains the usual vacuum and
matter contributions present at finite temperature, these last weighed
with the statistical distributions corresponding to the
boson and fermion internal lines in the self-energy diagram. 

Having carried out the sum over frequencies, it is now possible to
perform the analytical continuation of the function $I(i\omega, {\bf
p})$ to real energy values. We consider the analytical continuation to
the retarded function $I^r(p_0, {\bf p})$. Although the function
$I^r(p_0, {\bf p})$ contains in principle an imaginary as well as a
real part, in order to make the connection with the analysis in
Sec.~\ref{III}, here we concentrate on the real part. As stressed in
Ref.~\cite{Fueki}, attention must be paid to the imaginary part
when considering the complete tensor structure of the fermion
self-energy and/or a more complete form of the gauge boson propagator.
The real part of the function $I^r$ is defined as  
\be
   {\mbox{Re}}I^r(p_0, {\bf p})=\frac{1}{2}
   \left[I(i\omega\rightarrow p_0 + i\epsilon) +
   I(i\omega\rightarrow p_0 - i\epsilon)\right]\nonumber\\
   \label{RealI}
\ee 
and written explicitly as
\be
   {\mbox{Re}}I^r(p_0, {\bf p})&=&-\frac{1}{4E_1E_2}\Big\{
   \left[ 1 + f(E_2) - \tilde{f}(E_1) \right]\nonumber\\
   \!&\times&\! 
   {\cal P}\left( \frac{1}{p_0 - E_1 - E_2} - \frac{1}{p_0 + E_1 +
   E_2}\right)\nonumber\\
   \!&+&\! \left[ f(E_2) + \tilde{f}(E_1) \right]\nonumber\\
   &\times&
   {\cal P}\left( \frac{1}{p_0 - E_1 + E_2} - \frac{1}{p_0 + E_1 -
   E_2}\right)\Big\}\! \,,\nonumber\\
   \label{retar}
\ee
where ${\cal P}$ denotes principal part. It is now easy to see that
the integral in Eq.~(\ref{massfunceucl}) can now be expressed in
Minkowski space in terms of the functions 
\be
   I_1(p_0,p)&=& {\cal P} \int\frac{d^3k}{(2\pi)^3}
   \frac{{M}(k)}{4E_1E_2}\nonumber\\
   &&\left(\frac{1}{p_0 + E_1 + E_2} -
   \frac{1}{p_0 - E_1 - E_2}\right)\nonumber\\
   I_2(p_0,p)&=& {\cal P} \int\frac{d^3k}{(2\pi)^3}
   \frac{{M}(k)}{4E_1E_2}f(E_2)\Big\{\nonumber\\
   &&\left(\frac{1}{p_0 + E_1 + E_2} -
   \frac{1}{p_0 - E_1 - E_2}\right)\nonumber\\
   &+& 
   \left(\frac{1}{p_0 - E_1 + E_2} -
   \frac{1}{p_0 + E_1 - E_2}\right)\Big\}\nonumber\\
   I_3(p_0,p)&=& {\cal P} \int\frac{d^3k}{(2\pi)^3}
   \frac{{M}(k)}{4E_1E_2}\tilde{f}(E_1)\Big\{\nonumber\\
   &&\left(\frac{1}{p_0 - E_1 - E_2} -
   \frac{1}{p_0 + E_1 + E_2}\right)\nonumber\\
   &+& 
   \left(\frac{1}{p_0 + E_1 - E_2} -
   \frac{1}{p_0 - E_1 + E_2}\right)\Big\}\, ,\nonumber\\
   \label{I's}
\ee
which are thus written to separate the different vacuum and matter
contributions.
The choice equivalent to working with the lowest value of $i\omega$
is to take $p_0=0$. Also, since we are interested in looking for
critical behavior, we neglect ${M}^2$ for which $E_1\rightarrow k$. 

As a second simplification, let us ignore the angular dependence of
both the distribution functions or the mass function. This can be
thought of as considering the situation where $p\approx 0$. This is
the region of interest for the present analysis since, as shown in the
previous section, the small $p$ region is the relevant domain for the
development of a non-trivial solution for ${M}$. The above 
simplification allows to analytically perform the angular
integrations. Therefore, keeping in mind that we are interested in the
situation where $p\rightarrow 0$, the set of Eqs.~(\ref{I's}) becomes
\be
   I_1(p)&=&\frac{1}{8\pi^2 p}\int_0^\infty dk {M}(k) 
   \ln \left|\frac{p+2k}{p-2k}\right|\nonumber\\
   I_2(p)&=&\frac{1}{8\pi^2 p}\int_0^\infty dk {M}(k)f(k) 
   \ln \left|\frac{p+2k}{p-2k}\right|\nonumber\\
   I_3(p)&=&-\frac{1}{8\pi^2 p}\int_0^\infty dk {M}(k)\tilde{f}(k) 
   \ln \left|\frac{p+2k}{p-2k}\right|\, .
   \label{I'safterang}
\ee
Thus, in this limit, Eq.~(\ref{massfunceucl}) can be written in
Minkowski space as
\be
   {M}(p) &=& \frac{(3+\xi)\alpha}{2\pi p}
   \int_0^\infty dk{M}(k)\nonumber\\
   &&\left[ 1 + f(k) - \tilde{f}(k) \right]
   \ln \left|\frac{p+2k}{p-2k}\right|\, .
   \label{Mafterang}
\ee
We now use Eq.~(\ref{logapprox}) to write Eq.~(\ref{Mafterang}) as
\be
   {M}(p) &=& \frac{(3+\xi)\alpha}{2\pi p}\Big\{\nonumber\\
   &&
   \int_0^{p/2} dk{M}(k)\left[ 1 + f(k) - \tilde{f}(k) \right]
   \frac{4k}{p}\nonumber\\
   &+&
   \int_{p/2}^\infty dk{M}(k)\left[ 1 + f(k) - \tilde{f}(k)
   \right]\frac{p}{k}\Big\}\, .
   \label{Mafterapprox}  
\ee
Deriving Eq.~(\ref{Mafterapprox}) and multiplying by $p^3$, we obtain
\be
   p^3\frac{d{M}(p)}{dp}\!&=&\!-\frac{4(3+\xi)\alpha}{\pi}\nonumber\\
   \!&\times&\!
   \int_0^{p/2}dk{M}(k)\left[ 1 + f(k) - \tilde{f}(k)
   \right]k.
   \label{Mder}
\ee
For $p\rightarrow 0$, the integration region in Eq.~(\ref{Mder}) is
that of small $k$. Notice that, in spite of the presence of the
Bose-Einstein distribution, the integrand is infrared safe since
\be
   \left[ 1 + f(k) -
   \tilde{f}(k)\right]k\rightarrow T
   \label{limit}
\ee
when $k\rightarrow 0$. We can incorporate this behavior by
considering the dominant contribution to the integrand, namely, the one
coming from the low $k$ expansion of the function $f(k)\approx T/k$
writing
\be
   p^3\frac{d{M}(p)}{dp}=-\frac{4(3+\xi)\alpha}{\pi}
   T\int_T^{p/2}dk {M}(k)\, .
   \label{lowk}
\ee
This requires the mass function ${M}$ to satisfy
\be
   \left. p^3\frac{d{M}(p)}{dp}\right|_{p=2T}=0\, ,
   \label{boundary}
\ee
and, by further deriving Eq.~(\ref{lowk}), to also satisfy
the differential equation
\be
   p^3\frac{d^2{M}(p)}{dp^2} + 3p^2\frac{d{M}(p)}{dp}
   + \frac{2(3+\xi)\alpha^\prime}{\pi}{M}(p) = 0\, .
   \label{diffeq2}
\ee
Equation~(\ref{diffeq2}) becomes identical to Eq.~(\ref{diflog})
after the substitution $T\rightarrow 2T$. Thus, Eqs.~(\ref{diflog})
and~(\ref{diffeq2}) show that the scale for the onset of critical
behavior can be obtained by considering only the lowest Matsubara
mode in the analysis. 

\section{Conclusions}\label{concl}

We have studied the SDE for the fermion propagator to find non-trivial
solution for the mass function in qQED under the assumption that the
fermion-photon vertex is bare, the full fermion propagator can
be approximated by Eq.~(\ref{fullprop}) and that only the zeroth
frequency mode contributes dominantly. In order to stress the
relation to the case of QED3, we have performed the
analysis taking as independent parameters $\alpha'=\alpha T$ and $T$.
As expected, we observe that chiral symmetry is dynamically
broken above a critical value of coupling 
$\alpha'$ for a fixed temperature and is restored above a critical value
of temperature while we hold the coupling constant. We also
find that under simplifying assumptions, the result remains
qualitatively unchanged by summing over 
all the frequency modes. Although not explicitly worked out here,
the analysis can be also made considering $\alpha$ and $T$ as
the independent parameters. 

In order to clarify the meaning of the mass function found in this
work, recall that in vacuum, the ultraviolet behavior of the theory
is connected to the dynamical symmetry breaking phenomenon. The
connection, in the case of theories with anomalies (e.g. the
chiral anomaly) comes, on general grounds, through the impossibility
of simultaneously regularizing the theory while preserving the
classical symmetries. In fact, it is well known that in
qQED in 4 dimensions at $T=0$, the generated mass function is
proportional to the ultraviolet cut-off. However, the important point
to keep in mind is that this happens because this cut-off is the only
mass scale available in the theory. 
    
The former does not mean that the ultraviolet divergences of the
theory are essential for dynamical symmetry breaking. An example is
provided by QED3 at $T=0$, where chiral symmetry is dynamically broken
despite the fact that the theory is ultraviolet safe and
super-renormalizable. In this case, the mass scale available is
the dimensionful coupling. 

The above remarks apply to the case of the theory in vacuum and
keep being true at finite $T$ since the vacuum properties are left
untouched by the temperature.  
    
We should stress however that the mass function found in this work is
thermal, i.e. it vanishes for $T=0$. This means that, unlike
the vacuum case, the ultraviolet sector of the theory is not the one that
provides the mass scale for dynamical symmetry breaking. This can be
understood by noticing that any $T=0$ ultraviolet divergence is not
present at finite $T$ since the thermal distributions cut off the
integrals at large momentum. Thus, the mass scale is provided by the
temperature and in fact our results show, for the case of
qQED that we consider, that the generated mass function is
proportional to $T$. 

Criticality arises due to the existence of a dimensionless
parameter. In QED4 at $T=0$, it is $\alpha$. In contrast, 
in QED3 at $T=0$, there is no such parameter and therefore, chiral
symmetry is dynamically broken for all values of the coupling. In
QED4 at finite $T$, taking $\alpha' = \alpha T$ to be an independent
parameter, the dimensionless parameter is provided by $\alpha'/T$.
Given the absence of ultraviolet divergences when considering only
thermal effects, no additional scales come into play from the
ultraviolet sector. The scaling law found is thus unaffected by the
ultraviolet sector.

It will be interesting to see the impact on
results if we include the wave function renormalization $F$ and the
function $b$ in which case the analysis should account for the
imaginary part of the self-energy~\cite{Fueki}. It is also
important to repeat the calculation accounting for a more
complete fermion-photon vertex, studying the impact of this
treatment on the gauge dependence of the critical values for
chiral symmetry restoration. All this is for future. 

\section*{Acknowledgments}

AB gratefully acknowledges useful discussions with M.R. Pennington and
the hospitality extended to him by 
the Institute of Particle Physics Phenomenology (IPPP), University 
of Durham, and the National Centre of Physics,
Quaid-i-Azam University, during his visit there in the summer of 2002.
Support for this work has been received in part by CIC under grant
number 4.12, PAPIIT under grant number IN108001 and CONACyT under
grants number 32395-E, 32279-E, 35792-E and 40025-F.

\end{document}